\documentclass[conference]{IEEEtran}

\usepackage{ifpdf}
\usepackage{cite}
\ifCLASSINFOpdf
  \usepackage[pdftex]{graphicx}
\else
\fi
\usepackage{graphicx}
\usepackage{subfigure}
\usepackage{amsmath}
\usepackage{amssymb}
\usepackage{amsfonts}
\usepackage{amsmath,bm}
\usepackage{algorithmic}
\usepackage{algorithm}
\usepackage{array}
\usepackage{graphicx}
\usepackage{mathrsfs}
\usepackage{multirow, listings}
  \usepackage{verbatim}
\graphicspath{IEEEtran}

\usepackage{subfigure,cleveref}
\usepackage[letterpaper, top=0.75in, bottom=1in, left=0.625in, right=0.625in]{geometry}

\begin{document}
\renewcommand{\algorithmicrequire}{\textbf{Input:}}
\renewcommand{\algorithmicensure}{\textbf{Output:}}

\title{title}
\title{Joint Subcarrier and Power Allocation for Multi-User OFDM ISAC Systems}

\author{\IEEEauthorblockN{Jiayue Huang, Ning Wei, Min Qiu, and Aimin Tang}
\IEEEauthorblockA{Shanghai Jiao Tong University, Shanghai, China\\
Email: {\{huangjiayue, wei-ning, min\_qiu, tangaiming\}@sjtu.edu.cn}}
}

\maketitle


\begin{abstract}
Integrated sensing and communication (ISAC) is envisioned as a key enabling technology for 6G networks. Efficient resource sharing between communication and sensing is a fundamental challenge in orthogonal frequency-division multiplexing (OFDM) ISAC systems, particularly in multi-user scenarios, where channel diversity can be exploited to improve resource utilization. 
This paper investigates joint subcarrier assignment and power allocation for a downlink multi-user OFDM ISAC system, with the objective of minimizing the delay Cramér-Rao lower bound (CRLB) subject to constraints on the normalized integrated sidelobe level (ISL) over the region of interest (RoI), per-user data rates, and total transmit power. 
We first derive a convex formulation for sensing power allocation given fixed subcarrier assignments, and then develop an equivalent mixed-integer conic formulation of the joint optimization problem. To obtain computationally tractable solutions, we propose a conic relaxation-guided fix-and-optimize (CRG-FO) algorithm that integrates continuous relaxation, power re-optimization under fixed subcarrier assignments, and restricted neighborhood search. 
Simulation results demonstrate that joint subcarrier and power allocation reduces the distance CRLB relative to uniform subcarrier and power allocation across the tested ISL thresholds, rate loads, and user configurations. Power optimization on uniform sensing supports captures most of this gain, while adaptive subcarrier assignment provides only modest additional improvement. 
\end{abstract}

%
\IEEEpeerreviewmaketitle

\section{Introduction}

Integrated sensing and communication (ISAC) has been envisioned as a key technology for 6G wireless networks, enabling communication and sensing through shared hardware and spectrum resources \cite{Liu2022ISAC,Tang2024FDISAC}. 
To support sensing and communication within a unified system, dual-functional waveform design is essential for the different requirements of the two functions \cite{Wei2023Survey}. Among candidate waveforms, OFDM is a promising waveform for ISAC owing to its compatibility with existing communication systems and flexibility in resource allocation. 
Other multicarrier modulation schemes have also been investigated for ISAC \cite{Wei2023Survey}. In OFDM-based ISAC systems, reference signals (RSs) can be embedded in the time-frequency resource grid for sensing, while the remaining resource elements are allocated to communication data transmission. Accordingly, the placement and power allocation of sensing RSs need to be jointly designed to balance sensing performance and communication requirements.

Extensive research has been devoted to reference signal design for OFDM-based ISAC systems. Radar sensing using positioning reference signals (PRSs) is studied in \cite{Wei2023PRS}, where range and velocity estimation performance is analyzed. Multi-target detection and joint velocity, angle, and range estimation using synchronization signal blocks (SSBs) are developed in \cite{Awad2025SSB,Tang2026SSB}. 
Beyond reusing existing signals, a scattered RS design with joint optimization of RS spacing and power allocation is investigated in \cite{Zhao2023RS} to minimize the total transmit power subject to communication and sensing requirements.
In these designs, sensing RS placement is generally restricted to uniform or predefined patterns, motivating further investigation of more flexible resource allocation between sensing and communications.

To enable more flexible use of sensing and communication resources, adaptive resource allocation is investigated in \cite{Li2025Resource} to balance sensing resolution, peak sidelobe level, and communication sum-rate. Range-Doppler integrated sidelobe level (ISL) minimization under multi-user communication quality-of-service constraints is also studied through symbol-level precoding in \cite{Li2025Sidelobe}
, with the optimization focusing on sidelobe suppression rather than explicitly incorporating a CRLB-based estimation metric.
For orthogonal resource allocation between sensing and communication, a cross-domain OFDM waveform design is developed in \cite{Zhang2024CrossDomain}, where communication-centric and sensing-centric strategies are formulated by optimizing resource allocation and shaping the ambiguity function within a prescribed region of interest (RoI). The formulation in \cite{Zhang2024CrossDomain} considers a single communication user. 
Recent studies have investigated joint subcarrier and power allocation for multi-user OFDM-ISAC systems \cite{zheng2026qos}, as well as joint beamforming and subcarrier allocation \cite{hatami2026beamforming,huo2026joint}. These studies explore different aspects of joint resource optimization, with an emphasis on radar SNR/SINR or sensing-rate performance. 
More closely related to our work, \cite{yang2025joint} jointly optimizes subcarrier assignment and sensing/communication power allocation to maximize communication data rate, subject to delay-estimation CRLB, peak-sidelobe ratio, and total transmit-power constraints. 
Although these studies address closely related problems, the particular combination of minimizing the delay-estimation CRLB under a normalized RoI-ISL constraint and individual user-rate requirements has not been directly addressed. In particular, jointly optimizing sensing and communication resources under these requirements calls for a design that accounts for both sensing-subcarrier selection and power allocation, while ensuring the rate requirements of individual users.



Motivated by the above considerations, we investigate joint subcarrier assignment and power allocation for a downlink multi-user OFDM ISAC system in this paper. The design is formulated as a mixed-integer nonlinear programming (MINLP) problem that minimizes the delay CRLB subject to normalized RoI-ISL, total transmit power, and per-user minimum-rate constraints. The binary subcarrier assignments and their coupling with continuous power variables make the problem nonconvex. For a fixed sensing subcarrier set, we show that the sensing power budget is fully utilized at optimality and reformulate the resulting delay-CRLB minimization as a convex quadratically constrained quadratic program (QCQP). For the joint optimization, we develop an equivalent mixed-integer conic reformulation and propose a conic relaxation-guided fix-and-optimize algorithm. The algorithm combines continuous conic relaxation, validated power recovery for fixed assignments, and restricted mixed-integer neighborhood search to obtain a feasible solution. Simulation results show that the proposed scheme reduces the distance CRLB relative to uniform-allocation baselines. Further improvements over optimized sensing power on fixed uniform subcarriers are modest in the tested settings. The influence of user count is also evaluated under a fixed aggregate rate requirement.

\section{System Model}
\label{sec:system_model}

We consider a downlink multi-user OFDM ISAC system with a single-input, single-output (SISO) configuration. A BS serves \(U\) users simultaneously while performing monostatic sensing on a single target. The BS is assumed to support simultaneous transmission and reception, with self-interference neglected after sufficient isolation and cancellation \cite{tang2014balanced,bharadia2013full}. The system employs an OFDM waveform with \(N_c\) subcarriers. The subcarrier spacing is \(\Delta f\), and the OFDM symbol duration is \(T = 1/\Delta f\). The cyclic-prefix (CP) duration is denoted by \(T_{\mathrm{CP}}\), and the total OFDM symbol duration including the CP is \(T_\mathrm{sys}=T+T_\mathrm{CP}\). 
Each subcarrier is assigned exclusively to either sensing or communication with one of the \(U\) users, where the number of users \(U\) satisfies \(U\le N_c-1\).

\subsection{Transmit Signal Model}
Let \(\boldsymbol{\lambda} \in \{0,1\}^{N_c}\) and  \(\mathbf{b}_u \in \{0,1\}^{N_c}\) be the subcarrier assignment vectors, where \(\lambda_n=1\) indicates that subcarrier \(n\) is assigned to sensing, and \(b_{u,n}=1\) indicates that it is assigned to communication user \(u\). Their exclusivity is enforced by
\begin{equation}
    \sum_{u=1}^U b_{u,n} + \lambda_n \le 1, \quad \forall n.
\end{equation}

The complex symbol transmitted on the \(n\)-th subcarrier is denoted by \(s_n \in \mathbb{C}\), while \(p_{\mathrm{s},n}\geq 0\) and \(p_{\mathrm{c},u,n}\geq 0\) represent the sensing transmit power and the communication transmit power allocated to user \(u\) on subcarrier \(n\), respectively. Under the orthogonal allocation constraint, the effective transmit power on subcarrier \(n\) is given by
\begin{equation}
    |s_n|^2=\lambda_np_{s,n}+\sum_{u=1}^Ub_{u,n}p_{c,u,n}.
\end{equation}
The total transmit power satisfies \(\sum_{n=0}^{N_c-1}\left|s_n\right|^2\le P_t\).
The discrete-time baseband OFDM signal is given by
\begin{equation}\label{def:trans_sig}
    \bar{x}(\bar{n}) = \frac{1}{\sqrt{N_c}} \sum_{n=0}^{N_c-1} s_n e^{j2\pi (n-\lfloor N_c/2 \rfloor) \bar{n}/N_c}, \bar{n}=0,1,\ldots,N_c-1.
\end{equation}

\subsection{Multi-User Communication Model}

For each user \(u\), the discrete channel gain on subcarrier \(n\) is denoted by \(h_{u,n} \in \mathbb{C}\), which is assumed to be known at the BS. Assuming a frequency-selective channel with maximum delay shorter than \(T_\mathrm{CP}\) and negligible variation within each OFDM symbol, the received signal on subcarrier \(n\) assigned to user \(u\) after CP removal and FFT processing is
\begin{equation}
    y_{u,n} = b_{u,n} h_{u,n} s_n + w_{u,n},
\end{equation}
where \(w_{u,n} \sim \mathcal{CN}(0,\sigma_c^2)\) is the AWGN. The achievable data rate for user \(u\) is
\begin{equation}
    R_u = \frac{T}{T_\mathrm{sys}}\sum_{n=0}^{N_c-1} b_{u,n} \Delta f \log_2\!\left(1 + \frac{p_{\mathrm{c},u,n} |h_{u,n}|^2}{\sigma_c^2}\right).
\end{equation}
This rate expression is the communication performance metric, and we require that each user meets a minimum rate target \(R_u \ge \eta_u\).

\subsection{Sensing Signal Model and Performance Metrics}

For sensing, the BS transmits the sensing waveform and receives the echo reflected from the target. The received sensing signal on the \(n\)-th RE is modeled as
\begin{equation}
    y_{\mathrm{s},n} = A_{\mathrm{s}} \lambda_n \cdot s_n e^{-j2\pi f_n \tau} + w_{\mathrm{s},n},
\end{equation}
where \(A_{\mathrm{s}}\) is the complex two‐way attenuation constant (including radar cross section and path loss), \(\tau\) is the round‐trip time delay, and \(w_{\mathrm{s},n} \sim \mathcal{CN}(0,\sigma_{\mathrm{s}}^2)\) is the AWGN. We adopt a conditional delay CRLB in which \(A_{\mathrm{s}}\) is treated as fixed and known. Following the normalization in~\cite{liyanaarachchi2021optimized}, we set \(A_s=1\) and scale the noise variance accordingly. 
We focus on delay estimation for a target whose Doppler-induced inter-carrier interference can be neglected.

Two fundamental metrics are used to evaluate the sensing capability: the CRLB for delay that characterizes the estimation accuracy and the ambiguity function that characterizes the sidelobe structure. 
For the OFDM system, the resulting CRLB for delay \(\tau\) is given by~\cite{liyanaarachchi2021optimized}
\begin{equation}\label{eq:crlb_def}
    \mathrm{CRLB}(\hat{\tau}) = \frac{\sigma_s^2}{8\pi^2 \sum_{n=0}^{N_c-1} \lambda_n f_n^2 p_{\mathrm{s},n}},
\end{equation}
where \(\hat{\tau}\) denotes the estimator of \(\tau\) and \(f_n =(n-\lfloor N_c/2\rfloor)\Delta f\).
These expressions reveal that the sensing accuracy is directly influenced by the power allocation across subcarriers. 
Let \(\bar{x}_\mathrm{s}(\bar{n})\) denote the sensing component obtained from~(\ref{def:trans_sig}) by replacing \(s_n\) with \(\lambda_ns_n\).
Following the discrete-delay ambiguity function formulation in~\cite{Zhang2024CrossDomain}, we define
\begin{equation}\label{def:amb}
    \chi_a(\mu) 
    = \sum_{\bar{n}=0}^{N_c-1} \bar{x}_\mathrm{s}(\bar{n}) \bar{x}_\mathrm{s}^*(\bar{n}+\mu)
    =\sum_{n=0}^{N_c-1} \lambda_n p_{\mathrm{s},n} e^{-j2\pi \mu f_n/(N_c\Delta f)},
\end{equation}
where \(\mu\in\{0,1,\dots,N_c-1\}\) is the delay index.
The main lobe occurs at \(\mu=0\), for which \(\chi_a(0) =\sum_{n=0}^{N_c-1} \lambda_np_{\mathrm{s},n}\). 
The RoI is defined on this discrete delay grid as
\(
    \Omega_s = \left\{ 0, 1, \dots,  \frac{N_c}{b}-1 \right\}
\),
where \(b\) is a positive integer divisor of \(N_c\).
To quantify the sidelobe suppression within this RoI, we define the discrete ISL as
\begin{equation}\label{def:isl}
    \mathrm{ISL}_{\mathrm{RoI}} = \sum_{\mu \in \Omega_s} |\chi_a(\mu)|^2 - |\chi_a(0)|^2.
\end{equation}
The ISL serves as a key constraint in our optimization problems to control the aggregate sidelobe energy. We can use the normalized ISL to constrain the sidelobe levels, i.e., \(\mathrm{ISL}_{\mathrm{RoI}} / |\chi_a(0)|^2 \le \epsilon\), where \(\epsilon\) is a prescribed threshold. The ISL constraint limits sidelobe energy only on the integer delay grid, leaving off-grid ambiguity-function values unconstrained.

\section{Joint Resource Optimization}
\label{sec:algorithm}
To exploit the channel diversity gain for multi-user OFDM communications and enhance sensing performance, the resources of subcarriers and power are jointly optimized. The optimization aims to minimize the sensing CRLB while satisfying the communication rate requirements for each user and the sensing sidelobe constraint. Therefore, the problem is formulated as
\begin{subequations}\label{op:joint}
    \begin{align}
        \min_{\mathbf{p}_{\mathrm{s}}, \{\mathbf{p}_{\mathrm{c},u}\}, \boldsymbol{\lambda}, \{\mathbf{b}_u\}} \quad & \mathrm{CRLB}(\hat{\tau}) \\
        \text{s.t.} \quad & R_u \ge \eta_u, \quad \forall u,\\
        & \frac{\mathrm{ISL}_{\mathrm{RoI}}}{|\chi_a(0)|^2} \le \epsilon, \\
        & \lambda_n + \sum_{u=1}^U b_{u,n} \le 1, \quad \forall n, \\
        & \sum_{n=0}^{N_c-1} \left(\lambda_np_{\mathrm{s},n}+ \sum_{u=1}^Ub_{u,n}p_{\mathrm{c},u,n}\right) \le P_t, \\
        & \lambda_n, b_{u,n} \in \{0,1\}, \\
        & p_{\mathrm{s},n}\ge 0, \quad 0\le p_{\mathrm{c},u,n} \le P_{\mathrm{c,max}},
    \end{align}
\end{subequations}
where \(\mathbf{p}_\mathrm{s}\) and \(\mathbf{p}_{\mathrm{c},u}\in\mathbb{R}^{N_c} \) are power assignment vectors with entries \(p_{\mathrm{s},n}\) and \(p_{\mathrm{c},u,n}\) respectively,
and a per-subcarrier communication power limit \(P_\mathrm{c,max}\) is imposed to prevent excessive power concentration on individual subcarriers.

This is a mixed‐integer non‐convex program due to the discrete subcarrier assignments and the nonlinear coupling among assignment and power variables. Directly solving it is NP‐hard in general. We first examine sensing-only power allocation given a fixed subcarrier set, and then develop a CRG-FO algorithm for the joint problem.

\subsection{Sensing‐Only Optimization}
We first consider the sensing-only optimization problem over a fixed available subcarrier set \(\mathcal{A}\). 
To distinguish this subproblem from~(\ref{op:joint}) with a total power budget \(P_t\), let \(P_\mathrm{s,avail}>0\) denote the total power budget for sensing. 
Setting \(\lambda_n=1\) for \(n\in\mathcal{A}\) and zero otherwise, the problem is formulated as \begin{subequations}\label{op:sen_min_crlb}
    \begin{align}
        \min_{\mathbf{p}_{\mathrm{s}}} \quad & \mathrm{CRLB}(\hat{\tau}) \\
        \text{s.t.} \quad & \frac{\mathrm{ISL}_{\mathrm{RoI}}}{|\chi_a(0)|^2} \le \epsilon, \label{op:sen_min_crlb_ISL}\\ 
        & \sum_{n\in \mathcal{A}}  p_{\mathrm{s},n} \le P_{\mathrm{s,avail}}, \label{op:sen_min_crlb_pow}\\
        & p_{\mathrm{s},n} \geq 0, \quad \forall n \in\mathcal{A},\\
        & p_{\mathrm{s},n} = 0, \quad \forall n \notin \mathcal{A}.
    \end{align}
\end{subequations}
We first show that the power constraint~(\ref{op:sen_min_crlb_pow}) is active at optimality, so that the denominator in~(\ref{op:sen_min_crlb_ISL}) can be replaced by \(P_\mathrm{s,avail}^2\). To see this, suppose there exists an optimal solution \(\mathbf{p}_\mathrm{s}^*\) with total sensing power \(P_\mathrm{s}^*<P_{\mathrm{s},\text{avail}}\). Define a scaled power vector \(\tilde{\mathbf{p}}_\mathrm{s} = \alpha \mathbf{p}_\mathrm{s}^*\) with \(\alpha = P_{\mathrm{s},\text{avail}}/P_\mathrm{s}^* > 1\). According to~(\ref{def:amb}) and~(\ref{def:isl}), both \(\mathrm{ISL_{RoI}}\) and \(|\chi_a(0)|^2\) are multiplied by \(\alpha^2\), leaving the normalized ISL unchanged. Thus, \(\tilde{\mathbf{p}}_\mathrm{s} \) is also feasible. According to Eq.~(\ref{eq:crlb_def}), the scaled solution \(\tilde{\mathbf{p}}_\mathrm{s} \) yields a strictly lower CRLB, contradicting the optimality of \(\mathbf{p}_\mathrm{s}^*\). Therefore, any allocation using less than the available sensing power cannot be optimal, and every optimal solution satisfies \(\sum_{n\in\mathcal{A}}p_{\mathrm{s},n}^*=P_\mathrm{s,avail}\).

We next derive a quadratic representation of the ISL constraint. Because \(|\chi_a(\mu)|^2=\chi_a(\mu)\chi_a^*(\mu)\), it follows from~(\ref{def:isl}) that we can get
\begin{equation}
    \mathrm{ISL_{RoI}} 
    =\sum_{\mu=1}^{N_c/b-1}\sum_{n_1=0}^{N_c-1}\sum_{n_2=0}^{N_c-1}p_{\mathrm{s},n_1}p_{\mathrm{s},n_2}e^{-j2\pi\mu(n_1-n_2)/N_c}.
\end{equation}
Define matrix
\(\mathbf D\in\mathbb C^{N_c\times(N_c/b-1)}\) with entries \(D_{n,\mu}=e^{-j2\pi\mu(n-\lfloor N_c/2\rfloor)/N_c}\), for \(n=0,\ldots,N_c-1\) and \(\mu=1,\ldots,N_c/b-1\).
Then, ISL can be represented as
\begin{equation}
    \mathrm{ISL}_{\mathrm{RoI}}
    =\mathbf p_\mathrm{s}^\top\mathbf D\mathbf D^\mathsf{H}\mathbf p_\mathrm{s}
    =\|\mathbf D^\mathsf{H}\mathbf p_s\|_2^2,
\end{equation}
where \((\cdot)^\mathsf{H}\) denotes the Hermitian transpose, and \(\mathbf D\mathbf D^\mathsf{H}\) is a Hermitian positive semidefinite matrix. 

Using the full-power property,
\(|\chi_a(0)|^2=(\mathbf 1^\top\mathbf p_s)^2
=P_{s,\mathrm{avail}}^2\).
Consequently, the normalized ISL constraint~(\ref{op:sen_min_crlb_ISL}) becomes
\(
    \|\mathbf D^\mathsf{H}\mathbf p_s\|_2^2
    \leq\epsilon P_{s,\mathrm{avail}}^2.
\)

Finally, according to~(\ref{eq:crlb_def}), minimizing the CRLB is equivalent to maximizing the sum \(\sum f_n^2 p_{\mathrm{s},n}\).  The problem~(\ref{op:sen_min_crlb}) is therefore equivalent to the following convex QCQP
\begin{subequations} \label{op:sen_maxfp}
    \begin{align}
        \max_{\mathbf{p}_{\mathrm{s}}} \quad & \mathbf{f}_s^\top \mathbf{p}_{\mathrm{s}} \\
        \text{s.t.} \quad & \|\mathbf D^\mathsf{H}\mathbf p_s\|_2^2
    \leq\epsilon P_{s,\mathrm{avail}}^2, \\
        & \mathbf{1}^\top \mathbf{p}_{\mathrm{s}} = P_{\mathrm{s},\text{avail}}, \\
        & p_{\mathrm{s},n} \ge 0, \quad \forall n\in\mathcal{A},\\
        & p_{\mathrm{s},n} = 0, \quad \forall n \notin \mathcal{A},
    \end{align}
\end{subequations}
where \(\mathbf{f}_\mathrm{s} = (f_0^2,f_1^2,...,f_{N_c-1}^2)^\top\) and \(\mathbf{1}\in\mathbb{R}^{N_c}\) is the all-ones vector. This QCQP can be efficiently solved as a second-order-cone program using CVX with a conic solver such as MOSEK.

\subsection{Joint Sensing and Communication Optimization}
For the joint design, 
define the effective sensing power \(q_{\mathrm{s},n} = \lambda_np_{\mathrm{s},n}\) and effective communication power \( q_{\mathrm{c},u,n}= b_{u,n} p_{\mathrm{c},u,n}
\). Let \(\mathbf{q}_\mathrm{s}\) denote the sensing power vector with entries \(q_{\mathrm{s},n}\).
Since minimizing the delay CRLB is equivalent to maximizing the normalized sensing information, define
\begin{equation}
    \Phi(\mathbf{q}_{\mathrm{s}}) = \frac{1}{f_{\max}^2}\sum_{n=0}^{N_c-1}f_n^2q_{\mathrm{s},n}, \quad f_{\max} = \max_n\left|{f_n}\right|.
\end{equation}
With \(g_{u,n} = |h_{u,n}|^2/\sigma_c^2\), the binary dependence of the communication rate is represented by the closed logarithmic perspective as 
\begin{equation}
    r_{u,n} \leq\frac{T}{T_\mathrm{sys}} b_{u,n}\Delta f\log_2\!\left(1 + \frac{g_{u,n}q_{\mathrm{c},u,n} }{b_{u,n}}\right).
\end{equation}
Its exponential cone representation yields the following mixed-integer exponential-conic program (MIECP)
\begin{subequations}\label{op:miecp}
    \begin{align}
        \max_{\mathbf{q}_\mathrm{s},\{\mathbf{q}_{\mathrm{c},u}\},\mathbf{B},\{\mathbf{r}_u\}} & \Phi(\mathbf{q}_\mathrm{s})\\
        \text{s.t.} \quad
        & \sum_{n=0}^{N_c-1}r_{u,n}\geq \eta_u,\quad \forall u,\\
        & \left( b_{u,n} + g_{u,n}q_{\mathrm{c},u,n}, b_{u,n}, \frac{\ln 2}{1/T_\mathrm{sys}}r_{u,n} \right) \in \mathcal{K_{\exp}}, \\
        & \|\mathbf D^\mathsf{H}\mathbf q_\mathrm{s}\|_2 \leq \sqrt{\epsilon}\mathbf{1}^\top \mathbf{q}_\mathrm{s}, \\
        & \sum_{n=0}^{N_c-1}\left(q_{\mathrm{s},n}+\sum_{u=1}^Uq_{\mathrm{c},u,n}\right)\le P_t,\\
        &\sum_{u=1}^{U}b_{u,n}\le 1, \quad b_{u,n}\in\{0,1\}, \quad \forall u,n,\\
        &0\le q_{\mathrm{s},n}\le P_t\left(1-\sum_{u=1}^{U}b_{u,n}\right), \quad \forall n,\\
        &0 \le q_{\mathrm{c},u,n}\le P_{\mathrm{c,max}}b_{u,n}, \quad \forall u,n,
    \end{align}
\end{subequations}
where \(\mathbf{q}_{\mathrm{c},u}\) and \(\mathbf{r}_u\) are communication power and auxiliary rate vectors for user \(u\) with entries \(q_{\mathrm{c},u,n}\) and \(r_{u,n}\) respectively, \(\mathbf{B}\in\{0,1\}^{U\times N_c}\) is the assignment matrix with entries \(b_{u,n}\), and 
\[
    \begin{split}
        \mathcal K_{\exp}=\left\{(x,y,z):y>0, y\exp(z/y)\leq x\right\}\\
        \cup\left\{(x,0,z):x\geq0,z\leq0\right\}
    \end{split}
\]
is the closed exponential cone.
The proposed CRG-FO algorithm combines continuous conic relaxation, validated fixed-assignment power recovery, and relaxation-guided neighborhood search, providing an approximate solution to~(\ref{op:miecp}).

\subsubsection{Relaxation}
We first relax the binary assignment variables as \(0\le b_{u,n}\le1\). The resulting problem is a convex conic program. Solving this relaxation yields a generally fractional assignment \(\overline{\mathbf{B}}\), which is used to guide integer recovery and neighborhood construction. For each subcarrier, define the relaxed state weights as \(\pi_{0,n} = 1-\sum_{u=1}^{U}\overline b_{u,n}\) and \(\pi_{u,n} = \overline b_{u,n}, u=1,\ldots,U\).
Let state \(0\) denote a non-communication subcarrier. Since \(\pi_{0,n}+\sum_u\pi_{u,n} = 1\), these quantities form a soft assignment vector over the non-communication state and the \(U\) user states. These state weights are subsequently used for integer recovery and neighborhood selection.

\subsubsection{Fixed-Assignment Recovery}
For a binary assignment \(\mathbf{B}\), capped inverse water-filling over \(\mathcal{C}_u = \{n|b_{u,n}=1\}\) obtains the minimum communication power satisfying the rate and per-subcarrier power constraints.
An infeasible assignment is discarded. Any carrier receiving zero communication power is released to the non-communication state, yielding a cleaned binary assignment \(\widehat{\mathbf{B}}\). For a feasible communication allocation, define the total communication power, the remaining sensing-power budget, and the set of subcarriers available for sensing as
\(
    P_\mathrm{c}(\widehat{\mathbf{B}}) = \sum_{u=1}^U\sum_{n\in\mathcal{C}_u}\widehat q_{\mathrm{c},u,n}\),
\(
    P_\mathrm{s}(\widehat{\mathbf{B}}) = P_t-P_\mathrm{c}(\widehat{\mathbf{B}}) \),
and \(
    \mathcal{A}(\widehat{\mathbf{B}}) = \left\{n|\sum_{u=1}^U\widehat b_{u,n}=0\right\}\), respectively.
The sensing power allocation \(\widehat{\mathbf{q}}_\mathrm{s}\) and corresponding objective value \(\widehat\Phi=\Phi(\widehat{\mathbf{q}}_\mathrm{s})\) can be obtained by solving (\ref{op:sen_maxfp}) with \(\mathcal{A}=\mathcal{A}(\widehat{\mathbf{B}})\), \(P_{\mathrm{s,avail}} = P_\mathrm{s}(\widehat{\mathbf{B}})\). The recovered variables \(\widehat{\mathbf{B}}\), \(\{\widehat{\mathbf{q}}_{\mathrm{c},u}\}\), and \(\widehat{\mathbf{q}}_\mathrm{s}\) are then validated against all constraints of the original problem~(\ref{op:joint}). We collect the preceding power recovery and validation steps into a fixed-assignment evaluator, denoted by 
\begin{equation}\label{def:evaluator}
    \mathcal {E}(\mathbf{B}) = (\widehat{\mathbf{B}},
    \{\widehat{\mathbf{q}}_{\mathrm{c},u}\},
    \widehat{\mathbf{q}}_\mathrm{s},
    \widehat{\Phi}).
\end{equation}
If any step fails, we set \(\mathcal{E}(\mathbf{B})=\varnothing\).

Having defined how a fixed binary assignment is evaluated, we next construct the initial binary assignment candidates. The first candidate is generated by rounding \(\overline{\mathbf{B}}\), with each subcarrier assigned to the state \begin{equation}
    \ell_n^{\mathrm{rel}} = \underset{\nu\in\{0,1,\ldots,U\}}{\arg\max}\pi_{\nu,n},
\end{equation}
where \(\ell_n^{\mathrm{rel}}=0\) denotes the non-communication state and \(\ell_n^{\mathrm{rel}}=u\) assigns subcarrier \(n\) to user \(u\). For each user, the rounded carrier set is checked against its capacity under the per-subcarrier power limit. If this capacity is below \(\eta_u\), unassigned subcarriers are added in descending order of \(g_{u,n}\). The candidate is discarded only if the requirement remains infeasible after all available subcarriers have been examined. Total-power and sensing feasibility are subsequently checked by \(\mathcal E\). The candidate set also includes up to \(\min\{b,4\}\) uniform-comb assignments with different offsets, a communication-efficient assignment constructed with all subcarriers available for communication, and the assignment from the previous tighter ISL threshold. All candidates are evaluated by \(\mathcal E\), and the best feasible one is selected as the initial incumbent, denoted by \(\mathbf B^{\mathrm{inc}}\). 
If no candidate is feasible, a time-limited branch-and-bound search is attempted to seek an integer feasible solution. Its returned assignment is evaluated by \(\mathcal{E}\) for numerical verification. The recovered assignment initializes \(\mathbf{B}^\mathrm{inc}\). The algorithm terminates if no integer feasible solution is found or the evaluation fails.

\subsubsection{Relaxation-Guided Neighborhood Search}

Let \(\ell_n^{\mathrm{inc}}\) denote the state of subcarrier \(n\) in \(\mathbf{B}^\mathrm{inc}\), where \(\ell_n^\mathrm{inc}=u\) if its entry \(B^\mathrm{inc}_{u,n}=1\) and \(\ell_n^\mathrm{inc}=0\) otherwise. Subcarriers with \(\ell_n^{\mathrm{rel}}\neq\ell_n^{\mathrm{inc}}\) are ranked in descending order of \(\pi_{\ell_n^{\mathrm{rel}},n}-\pi_{\ell_n^{\mathrm{inc}},n}\). Then the remaining subcarriers are ranked in ascending order of \(\max_\nu\pi_{\nu,n}\). For ranking, ties in the maximum relaxed weight are resolved in favor of the incumbent state.

A neighborhood block consists of \(K\) consecutive subcarriers in this ordering. For each block, a time-limited MOSEK solver is used to solve~(\ref{op:miecp}) by re-optimizing the corresponding \(UK\) binary variables and continuous variables, while fixing all other binary variables to \(\mathbf{B}^\mathrm{inc}\).
Each returned assignment is passed through~(\ref{def:evaluator}) and accepted only if it is feasible and improves \(\Phi\). 
The ranking is updated after each block. Examining \(L\) successive blocks constitutes one sweep. The search stops when a sweep produces no improvement or \(S_{\max}\) sweeps have been completed. It then returns the validated incumbent. The CRG-FO procedure is summarized in Algorithm~\ref{alg:crgfo}.

\begin{algorithm}[t]
    \caption{Conic Relaxation-Guided Fix-and-Optimize}
    \label{alg:crgfo}
    \footnotesize
    \begin{algorithmic}[1]
        \REQUIRE \(U\), \(N_c\), \(b\), \(\epsilon\), \(P_t\), \(P_\mathrm{c,max}\), \(\mathbf{D}\), \(\{f_n\}\), \(\{g_{u,n}\}\), \(\{\eta_u\}\), \(K\), \(L\), \(S_{\max}\).
        \ENSURE A validated solution \(\mathcal{E}(\mathbf{B}^\mathrm{inc})\).
        \STATE Solve the relaxed problem~(\ref{op:miecp}) and get \(\overline{\mathbf{B}}\). Terminate if it is infeasible.
        \STATE Form \(\{\pi_{\nu,n}\}\) according to \(\overline{\mathbf{B}}\). 
        \STATE Evaluate different assignment candidates by~(\ref{def:evaluator}) , and retain the best feasible one as \(\mathbf{B}^\mathrm{inc}\). If no seed is feasible, obtain and validate an incumbent by a short full-MIECP search.
        \FOR{\(s=1,\ldots,S_{\max}\)}
            \STATE Set \(\mathrm{improved}\leftarrow\mathrm{false}\).
            \FOR{\(l=1,\ldots,L\)}
                \STATE Update the relaxation-guided ranking and solve the next \(K\)-subcarrier restricted MIECP; apply~(\ref{def:evaluator}) to its integer candidate.
                \STATE Upon a strict improvement, update the incumbent and set \(\mathrm{improved}\leftarrow\mathrm{true}\).
            \ENDFOR
            \STATE If \(\mathrm{improved}=\mathrm{false}\), break.
            \ENDFOR
        \STATE \textbf{return} \(\mathcal{E}(\mathbf{B}^\mathrm{inc})\).
    \end{algorithmic}
\end{algorithm}

\subsection{Computational Complexity Analysis}

\emph{1) Sensing-only optimization:} Let \(N_{\mathrm{s}}=|\mathcal{A}|\) and \(D_\mathrm{s}=N_{\mathrm{c}}/b-1\). 
Under a dense primal-dual interior-point method, its worst-case complexity is approximately \(\mathcal{O}\!\left(\sqrt{N_{\mathrm{s}}+2}\left(N_{\mathrm{s}}^3+D_\mathrm{s}N_{\mathrm{s}}^2\right)\log\frac{1}{\delta}\right)\), where \(\delta\) denotes the solution accuracy.

\emph{2) Joint optimization:} Let \(\mathcal{C}_{\mathrm{conic}}\) denote the cost of solving one continuous conic subproblem, and let \(\mathcal{C}_{E}\) denote the cost of fixed-assignment power recovery. At most \(C_\mathrm{seed}=\min\{b,4\}+3\) seeds are evaluated. The worst-case complexity of CRG-FO is bounded by 
\[\mathcal{O}\Bigg(\left[1+S_{\max}L2^{UK}\right]\mathcal{C}_{\mathrm{conic}}
+\left[C_\mathrm{seed}+S_{\max}LK\right]\mathcal{C}_{E}\Bigg).
\]
When \(K\ll N_{\mathrm{c}}\), the restricted searches reduce the worst-case search bound relative to the \(\mathcal{O}(2^{UN_{\mathrm{c}}}\mathcal{C}_{\mathrm{conic}})\) bound for the full MIECP.
\section{Simulation Results}\label{sec:sim}

\begin{figure*}[!t]
    \centering
    \begin{minipage}[t]{0.315\textwidth}
        \centering
        \includegraphics[width=\linewidth]{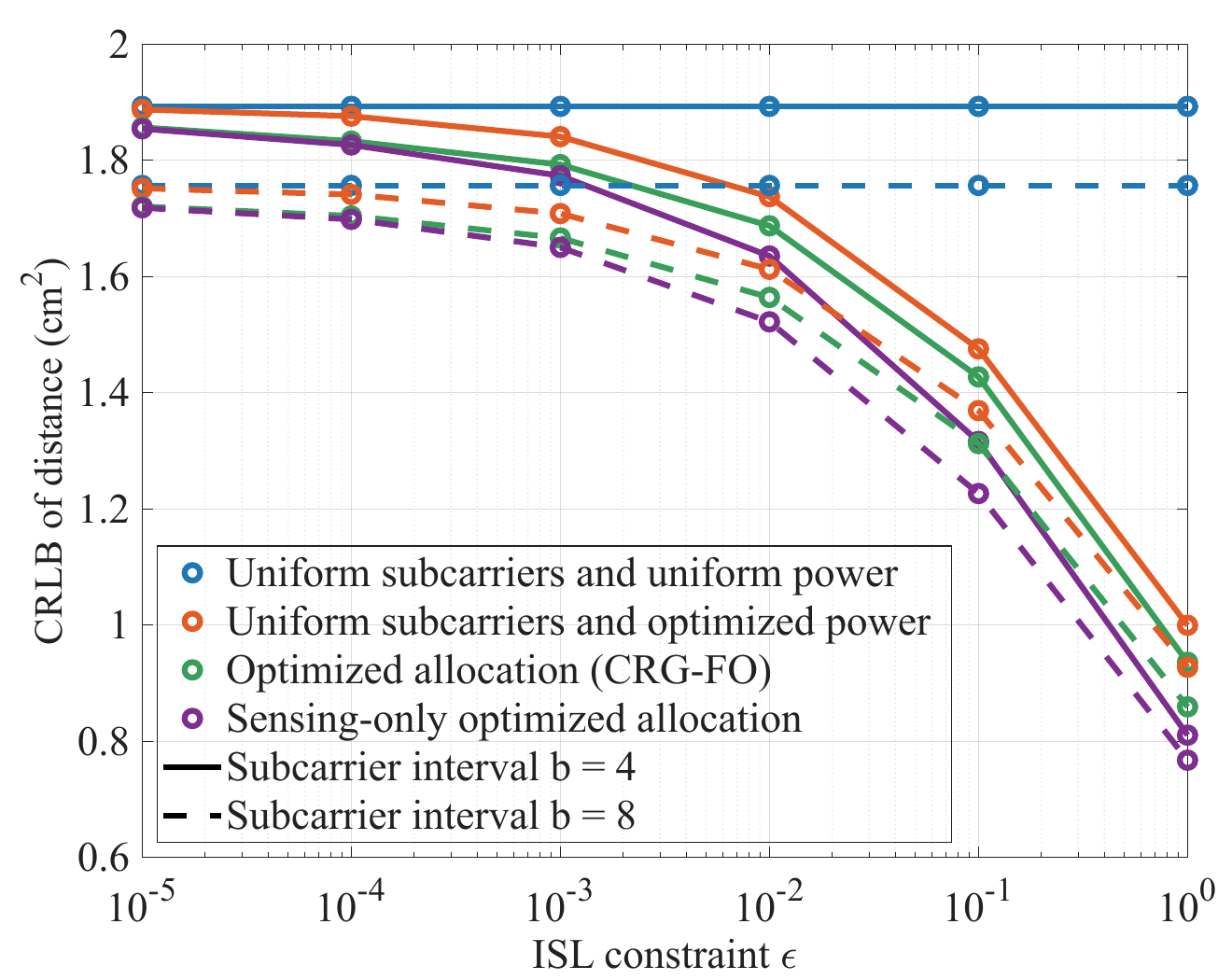}
        \caption{CRLB versus ISL constraint \(\epsilon\).}
        \label{fig:crlb_epsilon_roi}
    \end{minipage}\hfill
    \begin{minipage}[t]{0.32\textwidth}
        \centering
        \includegraphics[width=\linewidth]{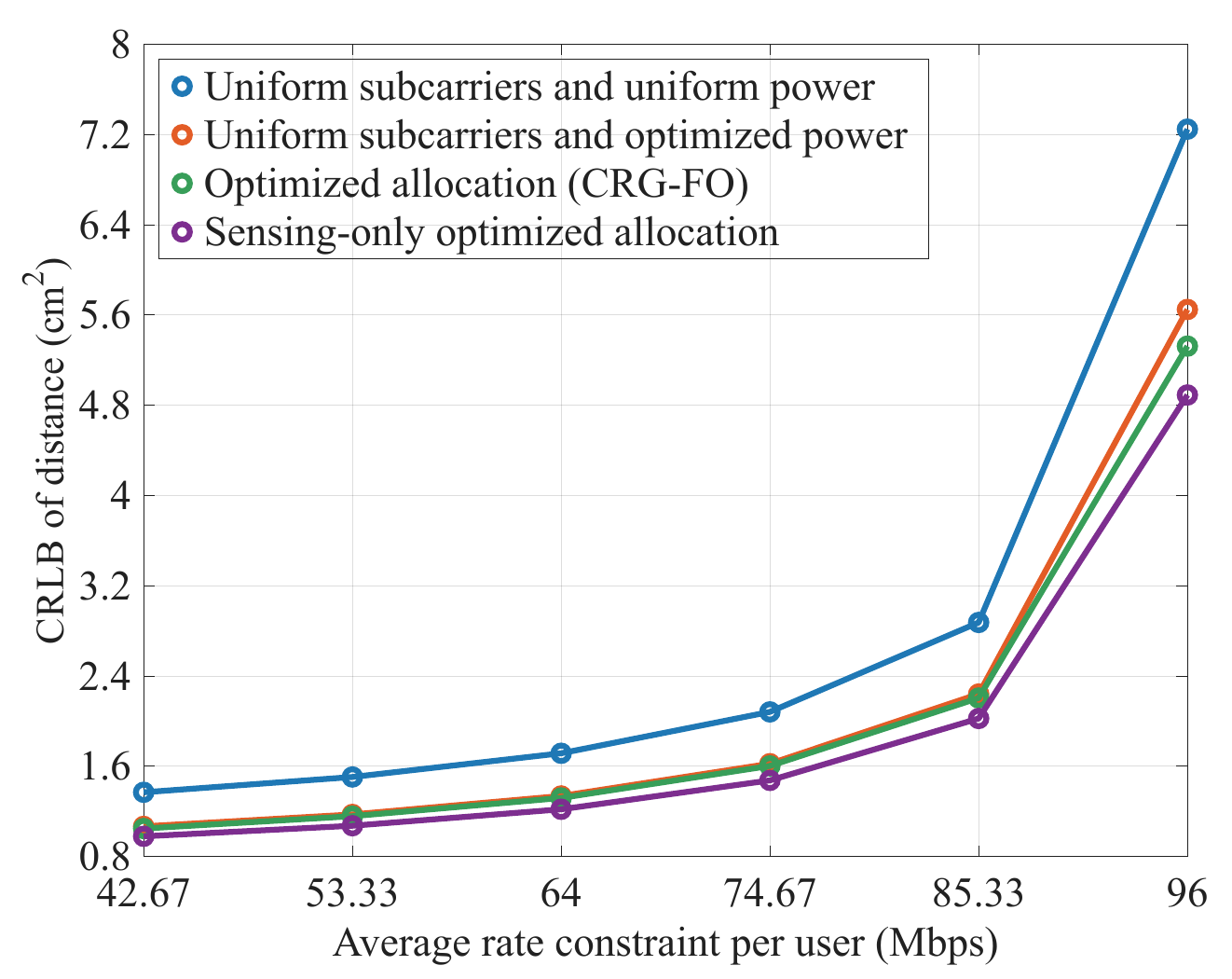}
        \caption{CRLB versus average-rate constraints.}
        \label{fig:crlb_averageRate}
    \end{minipage}\hfill
    \begin{minipage}[t]{0.32\textwidth}
        \centering
        \includegraphics[width=\linewidth]{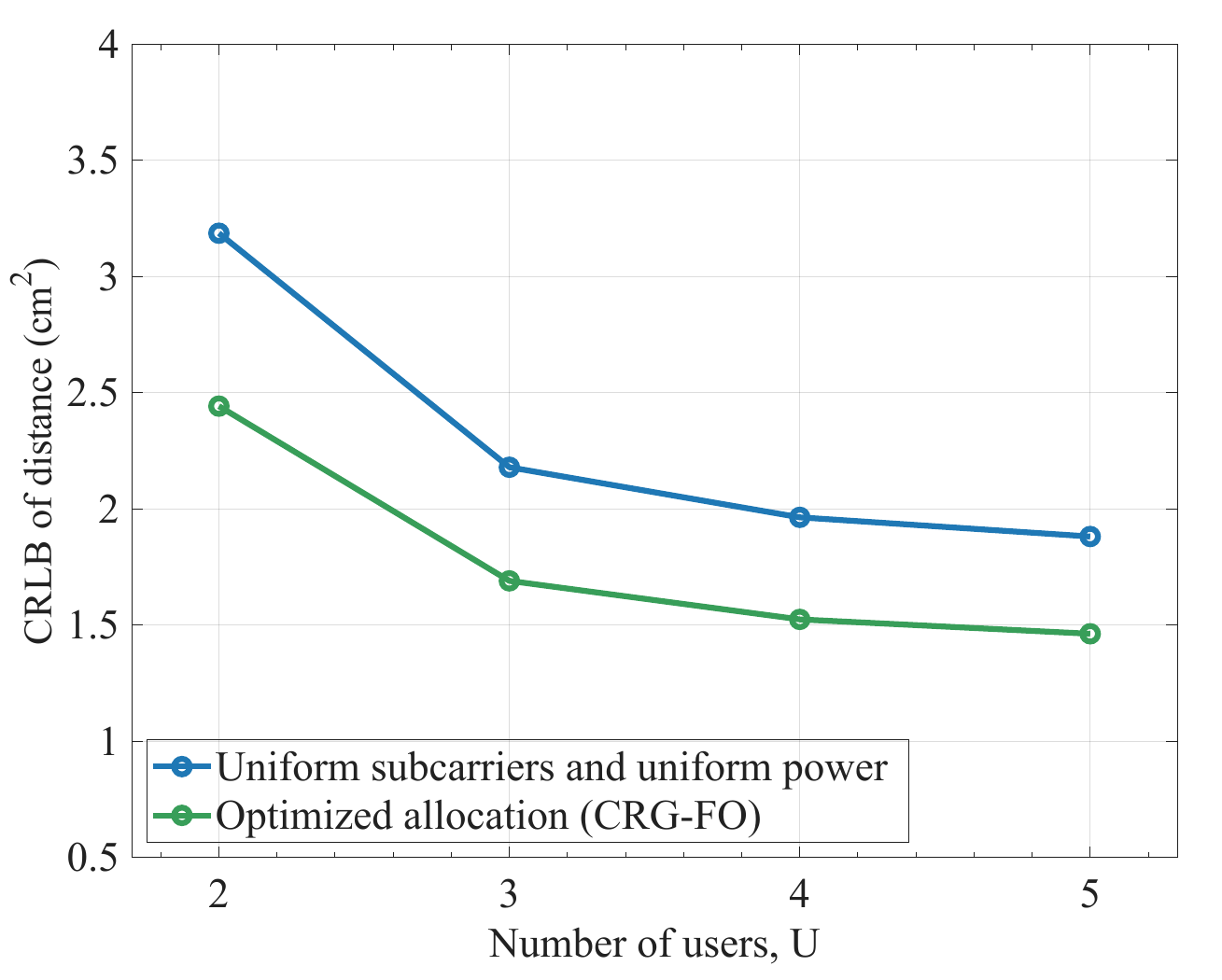}
        \caption{CRLB versus the numbers of users \(U\).}
        \label{fig:crlb_user}
    \end{minipage}
\end{figure*}

In this section, we evaluate the performance of the proposed joint sensing and multi-user communication resource allocation algorithm through numerical simulations. We consider an OFDM system with \(N_c = 512\) subcarriers and subcarrier spacing \(\Delta f = 240\) kHz, operating at carrier frequency \(f_c = 25\) GHz. The single OFDM symbol duration is \(T = 1/\Delta f\), with \(T_\mathrm{CP} = T/8\). The power parameters are set to \(P_t = 2\) W, \(P_{\mathrm{c,max}} = 0.1\) W, and \(\sigma_\mathrm{s}^2 = \sigma_c^2 = 10^{-3}\). We assume a simplified two-path frequency-selective Rayleigh channel. The two paths have independent zero-mean complex Gaussian gains with relative average powers of 0 and -5 dB, normalized to unit total average power, and independent delays are uniformly distributed within the CP. Results are averaged over common channel realizations. MOSEK is used with \(K=32\),\(L=2\) and \(S_\mathrm{max}=1\). To express the estimation accuracy in distance units, we convert the delay CRLB using \(d = \frac{c\tau}{2}\), yielding \(\mathrm{CRLB}(\hat{d}) = \frac{c^2}{4}\mathrm{CRLB}(\hat{\tau})\)) reported in \(\mathrm{cm}^2\).
Four allocation strategies are evaluated with the same total power limit \(P_t\) and are summarized as follows:
\begin{itemize}
    \item Uniform subcarriers and uniform power: Sensing subcarriers are reserved at spacing \(b\). Capped inverse water-filling minimizes the communication power \(P_\mathrm{c}\) for a feasible assignment over the remaining subcarriers, and the resulting \(P_\mathrm{s}=P_t-P_\mathrm{c}\) is uniformly allocated among the sensing subcarriers.
    \item Uniform subcarriers and optimized power: The same subcarrier assignments and \(P_\mathrm{s}\) are retained. The sensing power allocation is optimized according to~(\ref{op:sen_maxfp}).
    \item Optimized allocation (CRG-FO): The subcarrier assignments and power allocation are jointly optimized by Algorithm~\ref{alg:crgfo}.
    \item Sensing-only optimized allocation: The sensing power allocation is optimized using ~(\ref{op:sen_maxfp}), with all subcarriers available and the total sensing power set to that obtained by CRG-FO.
\end{itemize}

The effects of the sidelobe constraint and RoI are evaluated for two users with rate targets of 53.33 and 74.67 Mbps. The corresponding results for \(b=4\) and \(b=8\) are presented in Fig.~\ref{fig:crlb_epsilon_roi}. The optimized schemes achieve progressively lower CRLBs as \(\epsilon\) increases, whereas the uniform-power baselines remain constant because their fixed comb supports and equal sensing powers already satisfy the sampled RoI-ISL constraint. At \(\epsilon=1\), CRG-FO achieves approximately 0.936 and 0.859 \( \mathrm{cm}^2\) for \(b=4\) and \(b=8\), respectively. However, the additional reductions over optimized power on uniform supports are only \(6.4\%\) and \(7.4\%\). 
These modest additional gains arise because the uniform supports already span the full bandwidth and include subcarriers with large frequency offsets. Optimizing their powers therefore captures most of the improvement. CRG-FO provides further gains by adapting the sensing and communication supports, but requires mixed-integer searches. The sensing-only scheme ignores communication feasibility and is therefore an optimistic sensing reference rather than a lower bound on the joint problem. Overall, uniform subcarriers with optimized power provide an attractive low-complexity alternative in the tested setting.

To investigate the sensing cost of increasing communication demand, the average per-user rate target is varied from 42.67 to 96 Mbps with \(U=2\), \(b=4\), and \(\epsilon=10^{-1}\). 
As shown in Fig.~\ref{fig:crlb_averageRate}, all four CRLB curves increase with the rate requirement. 
Higher communication requirements restrict the feasible resource allocation and leave less power available for sensing under the fixed total budget. The sensing-only reference also deteriorates because its power budget follows the residual sensing power obtained by CRG-FO. 

The influence of the user count is evaluated for \(U\in\{2,3,4,5\}\) with \(b=4\) and a fixed aggregate rate of 128 Mbps. The per-user targets are equally spaced from \(128/U-10.67\) to \(128/U+10.67\) Mbps. As shown in Fig.~\ref{fig:crlb_user}, the average distance \(\mathrm{CRLB}\) decreases for both schemes. CRG-FO consistently outperforms the uniform subcarriers and uniform power allocation scheme. Under the fixed aggregate rate constraint, the decrease may partly reflect a user-diversity gain, since a larger user set provides more varied channel conditions for resource allocation. This can reduce the communication power required and leave more power available for sensing, while the lower per-user rate targets as \(U\) increases may also contribute to the improvement.
\section{Conclusion} \label{sec:conclusion}
This paper investigated joint subcarrier assignment and power allocation for a downlink multi-user OFDM-ISAC system. The delay CRLB was minimized subject to normalized RoI-ISL, per-user rate, and total power constraints. 
A convex formulation was derived for sensing power allocation on fixed subcarriers. The joint problem was then reformulated as a MIECP. The proposed CRG-FO algorithm combined continuous conic relaxation, fixed-assignment power recovery, and restricted neighborhood searches to obtain feasible allocations. 
Numerical results showed that CRG-FO reduced the distance CRLB relative to uniform subcarrier and power allocation. 
Power optimization on uniform sensing supports captured most of the overall gain, whereas adaptive subcarrier assignment provided further but generally modest improvements at the cost of mixed-integer searches. These findings highlight a performance-complexity tradeoff and identify uniform sensing supports with optimized power as an attractive low-complexity alternative in the considered setting. 

\bibliographystyle{IEEEtran}
\bibliography{reference}
\end{document}